\documentclass[conference]{IEEEtran}

\usepackage{amsmath,graphicx,verbatim,amssymb,amsfonts,comment}
\usepackage{epstopdf}
\usepackage[export]{adjustbox}
\usepackage{float,subfig}
\usepackage{array}
\usepackage{multirow}
\usepackage[ruled,vlined]{algorithm2e}
\usepackage{flushend}
\usepackage{xcolor}

\SetKw{step}{Step}{}
\SetKwProg{st}{Step}{}{}
\SetKwProg{sts}{Steps}{}{}
\SetKwProg{do}{Do}{}{}

\DeclareMathOperator*{\argmax}{arg\,max}
\DeclareMathOperator*{\argmin}{arg\,min}

\graphicspath{{./graphics/}}

\begin{document}

\title{Angle estimation using mmWave RSS  measurements with enhanced multipath information}

\author{%
  \IEEEauthorblockN{%
    Neharika Valecha\IEEEauthorrefmark{1}, 
    Jesus Omar Lacruz\IEEEauthorrefmark{2}, 
    Michael Lentmaier\IEEEauthorrefmark{1}, 
   Joerg Widmer\IEEEauthorrefmark{2}, and %
   Fredrik Tufvesson\IEEEauthorrefmark{1}
  }%
  \IEEEauthorblockA{\IEEEauthorrefmark{1} Department of Electrical and Information Technology, Lund University, Lund, Sweden}%
  \IEEEauthorblockA{\IEEEauthorrefmark{2} IMDEA Networks Institute, Madrid, Spain}%

\thanks{This project has received funding from the European Union’s EU Framework Programme for Research and Innovation Horizon 2020 under Grant
Agreement No. 861222.}%
}

\IEEEoverridecommandlockouts

\maketitle
\setlength{\textfloatsep}{10pt}
\IEEEpubidadjcol

\begin{abstract}
mmWave communication has come up as the unexplored spectrum for 5G services. With new standards for 5G NR positioning, more off-the-shelf platforms and algorithms are needed to perform indoor positioning. An object can be accurately positioned in a room either by using an angle and a delay estimate or two angle estimates or three delay estimates. We propose an algorithm to jointly estimate the angle of arrival (AoA) and angle of departure (AoD), based only on the received signal strength (RSS). We use mm-FLEX, an experimentation platform developed by IMDEA Networks Institute that can perform real-time signal processing for experimental validation of our proposed algorithm. Codebook-based beampatterns are used with a uniquely placed multi-antenna array setup to enhance the reception of multipath components and we obtain an AoA estimate per receiver thereby overcoming the line-of-sight (LoS) limitation of RSS-based localization systems. We further validate the results from measurements by emulating the setup with a simple ray-tracing approach.
\end{abstract}

\IEEEpeerreviewmaketitle


\section{Introduction}
\label{sec:introduction}
The ever-increasing need for larger bandwidths and higher data rates has led the telecom industry to move to hitherto unexplored higher frequency millimeter-wave (mmWave) band (20-300 GHz). This range of frequencies has quasi-optical properties that make it ideal for positioning applications. From the Internet of Things (IoT) to Vehicle-to-everything (V2X), most emerging applications need location awareness \cite{zafari2019survey}. In the past, Global Navigation Satellite Systems have been used for UE positioning, but that is preferable for outdoor systems. When trying to position UEs indoors, range-based or range-free algorithms are used.
 
 mmWave spectrum frequencies (over 24 GHz), due to their shorter wavelengths have longer, narrower beams \cite{shafi2018microwave} and are ideal for positioning. Consequently, the higher path loss due to the shorter wavelengths makes the use of massive antenna arrays mandatory. Furthermore, using beamforming in combination with a large number of antennas leads to highly directional beams. Fully digital beamforming, which supports massive MIMO arrays (upwards of $128 \times 256$) has emerged as a 'paradise for positioning'. With calibrated systems, it offers the use of super-resolution algorithms that can give position estimates with centimeter-level accuracy. However, the hardware and power constraints hinder its commercialization. Thus, hybrid beamforming is a solution, as it provides a trade-off between power consumption and position resolution \cite{shastri2022review}.

Localization using radio frequencies has its own challenges. 5G systems augment the use of massive antenna arrays which allow for increased resolvability in the angular domain. Most systems rely strongly on line-of-sight (LoS) information for positioning but these signal components are often blocked in indoor environments. Range-based localization hinges on time synchronization between the transmitter (TX) and receiver (RX) and requires calibrated antennas and coherent communication between TX and RX. Multipath components (MPCs) can be exploited for location information if effectively resolved \cite{meissner2014multipath}, \cite{meissner2012multipath}, \cite{leitinger2016cognitive}. Received Signal Strength (RSS) based localization is thus effective and can exploit hybrid beamforming setups for accurate positioning \cite{li2020rss}.
 
In this paper, we propose an RSS-based least squares (RSS-LS2D) algorithm to jointly estimate the Angle of Arrival (AoA) and the Angle of Departure (AoD) by minimizing over both TX and RX beampatterns. Our algorithm is an extension of the algorithm proposed in \cite{lacruz2020mm}, referred to as RSS-LS1D henceforth. It is 1D because either the AoA is estimated at the RX or the AoD at the TX, at one time. The algorithms are applied to channel measurements obtained using the mm-FLEX testbed described in \cite{lacruz2020mm}. As the RSS measurements rely on the strength of the signal, we use a special receiver with 4 RX antenna arrays placed such that they receive MPCs over a $360^\circ$ range. We show that our RSS-LS2D algorithm estimates AoA/AoD accurately even in the absence of LoS. Further, we validate our results by emulating the measurement environment using ray tracing.

\begin{figure*}[!t]
    \centering
    \includegraphics[width =\textwidth]{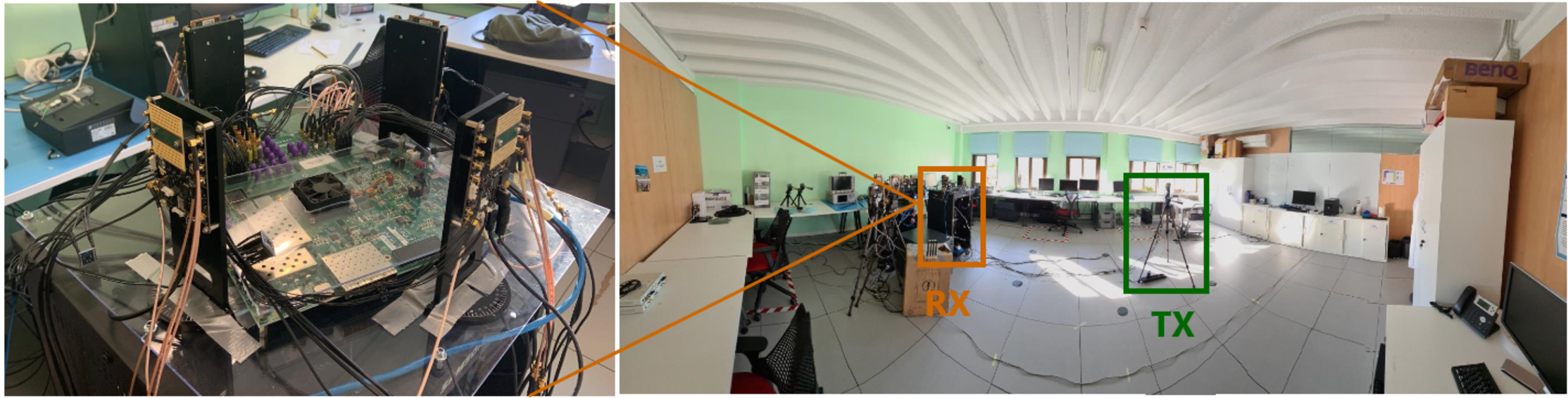}
    \caption{\textit{Left}: The 4 RX antenna arrays are connected to the SIVERS FPGA to perform real-time processing. \textit{Right}: Room at IMDEA with TX and RX placed at positions $\mathbf{p}_{1} = [0 \, \mathrm{m}, 2.4 \, \mathrm{m}]$ and $\mathbf{p}_{\mathrm{RX}} = [0 , 0]$, respectively.}
    \label{setup3}
\end{figure*}


\section{Setup}

\subsection{mm-FLEX Testbed}
\begin{figure}[t]
    \centering
    \vspace*{-1em}
    \includegraphics[width =0.8\linewidth]{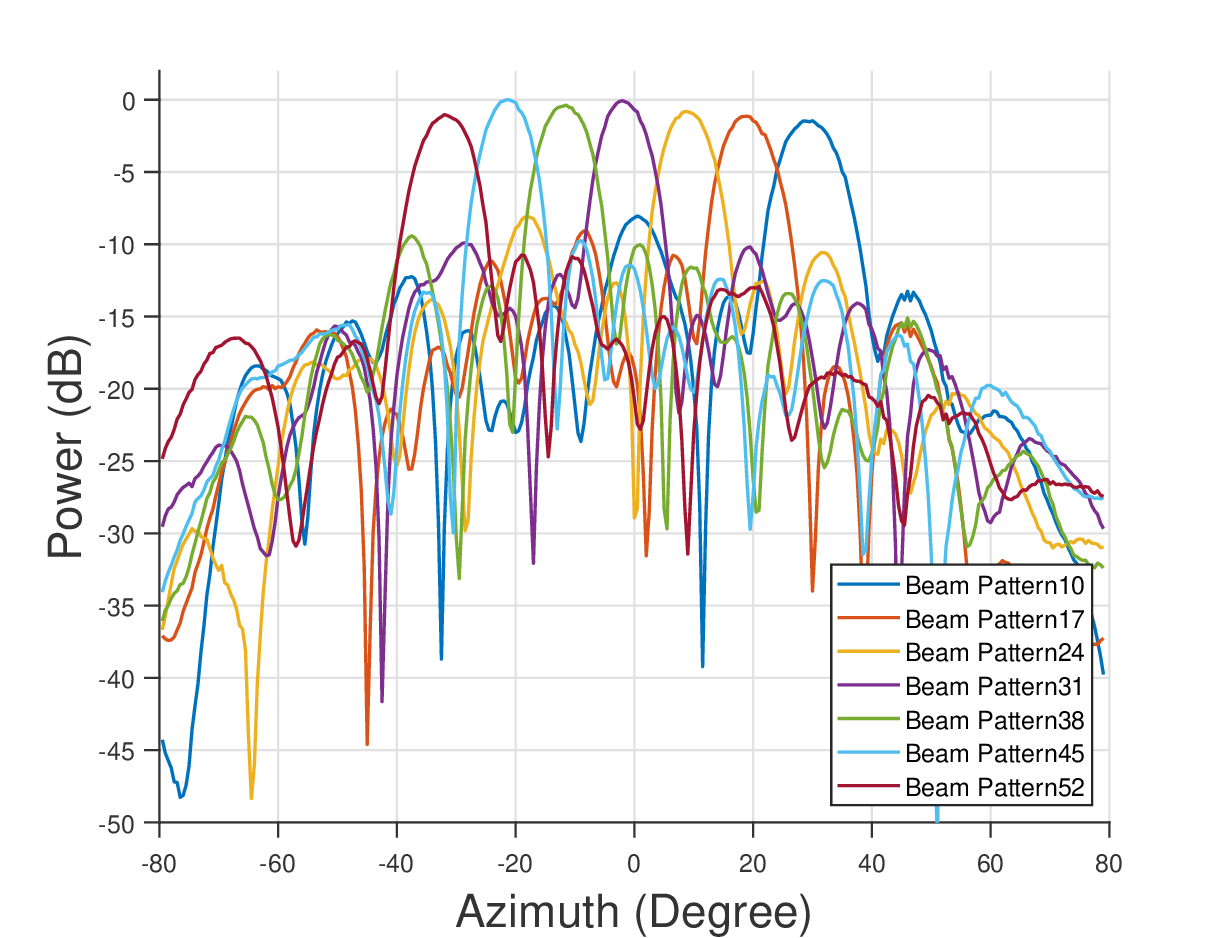}
    \caption{Illustration of a selection of measured beam patterns used in mm-FLEX.}
    \label{beampattern}
\end{figure}

 We use Millimeter Wave Mobile FulL-Bandwidth EXperimentation Platform (mm-FLEX), an open-source research platform with a 60~GHz RF front end that supports bandwidths up to 2~GHz. The TX and RX each have the AMC599 board, integrating a Xilinx Kintex Ultrascale FPGA, highspeed AD/DA converters, and DDR4 memory banks \cite{vada1} and an AMC726 board with an Intel Core i7 processor \cite{vada2}. It uses the EVK06002 development kit from Sivers IMA \cite{sivers1} with 60~GHz up/down converters as the RF front-end for mm-FLEX. The kit includes 2D analog beam-forming capabilities through phase shifters with a 6-bit resolution for each antenna element. In addition, the kit incorporates fast beam-switching capabilities through simple pulses via a GPIO interface, that allows switching beam patterns every $10 \, \mathrm{ns}$. However, the kit has a maximum RF settling time of $35 \, \mathrm{ns}$ when changing beam patterns. The RX setup is as shown in Fig.~\ref{setup3}. The TX uses a single RF linear antenna array with 16 antennas and the receiver has 4 separate arrays of the same type. The 4 RX arrays are placed orthogonally such that they receive multipath information and provide a quasi-omnidirectional view of the room. Each TX and RX array has 64 codebook-based beampatterns with main lobes at angles ranging from -45$^\circ$ to 45$^\circ$, illustrated in Fig.~\ref{beampattern}.
 The TX array and each of the RX arrays are loaded with the same codebook and have the same beampatterns, each covering an angular span of approximately $160^\circ$. In particular, $\mathrm{RX}_1$ has an angular span from $-79.5^\circ$ to $80^\circ$, $\mathrm{RX}_2$ has an angular span from $10.5^\circ$ to $170^\circ$, $\mathrm{RX}_3$ has an angular span from $100.5^\circ$ to $260^\circ$ and $\mathrm{RX}_4$ has an angular span from $190.5^\circ$ to $350^\circ$. Thus, all 4 RX arrays together cover a $360^\circ$ range with some overlap. The TX array has the same angular span as $\mathrm{RX}_3$.

 \subsection{Simulation and Measurement Environment}
\begin{figure}[t]
    \centering
    \includegraphics[width =\linewidth]{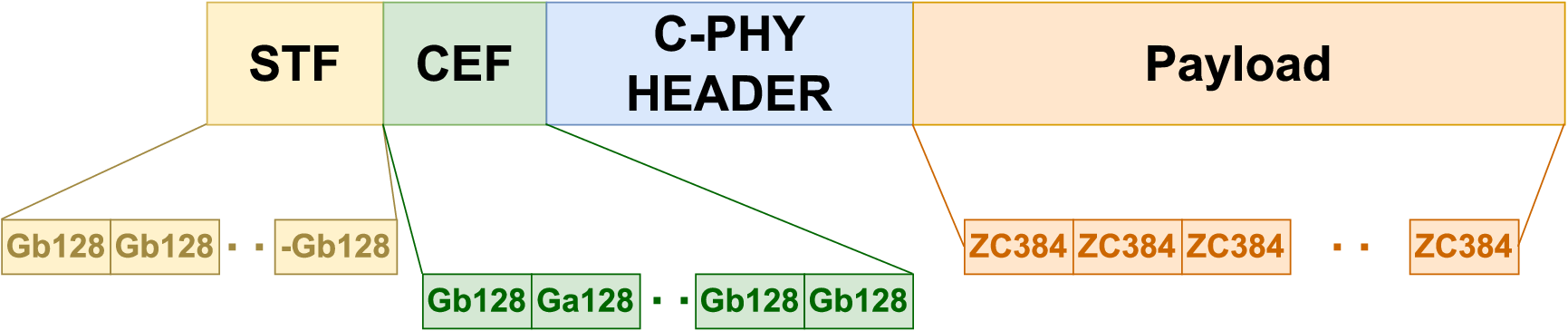}
    \caption{Structure of the transmitted frame with Golay sequences of length 128 bits and Zadoff-Chu sequences of length 384 bits.}
    \label{frame}
\end{figure}

We validate the proposed RSS-LS2D algorithm with real measured data as well as synthetically generated data. The measurements were carried out at IMDEA, the room with the TX and RX is shown in Fig.~\ref{setup3}. The room is $5 \, \mathrm{m} \times 8 \, \mathrm{m}$. As described in \cite{lacruz2020mm}, an IEEE 802.11 compatible frame is transmitted with Golay (Ga128) sequences in the preamble and Zadoff-Chu sequences as payload, having a bandwidth of 1.76 GHz. The hardware includes a square-root-raised-cosine (SRRC) filter which was switched off during our experiments. Instead, the frame was up/downsampled by using Fast Fourier Transform (FFT) as in 4G LTE systems. The frame structure can be seen in Fig.~\ref{frame}. 
The preamble consists of the Short Training Field (STF), Channel Estimation Field (CEF), and the C-PHY header containing information about the MCS, data length, and some other data. The STF contains 25 repeated Golay (Gb128) sequences of length 128 samples each and is needed for packet detection and synchronization. The CEF are four pairs of complementary Golay sequences. Zadoff-Chu sequences of length $K=384$ samples\footnote{ The size of the payload is limited by the storage available in the hardware. As the channel processing happens in real-time, the total data over all beampatterns should not exceed the DDR4 memory bank capacity.} are used as payload data, which can be generated as $$z_k= e^{-j \pi \frac{uk(k+c+2q)}{K}}$$ where $k = 0,1, \ldots, K-1$, $u= 0,1, \ldots, K-1$ and $\gcd(u, K)=1$, $c = \mathrm{mod}(K, 2)$ and $q \in \mathbb{Z}$. The frame is calibrated to receive the same Zadoff-Chu sequence over the $N_{\mathrm{BP},\mathrm{RX}}$ beampatterns. We obtain this frame at each of the four RX. The transmission and reception takes place over all $N_{\mathrm{BP},\mathrm{TX}} \times N_{\mathrm{BP},\mathrm{RX}}$ beampatterns, unlike in \cite{lacruz2020mm} where a beam refinement protocol (BRP) was used to find aligning beampatterns and the frame was transmitted over only the strongest TX beam pattern.

\begin{figure}[t]
    \centering
    \includegraphics[width =0.48\textwidth]{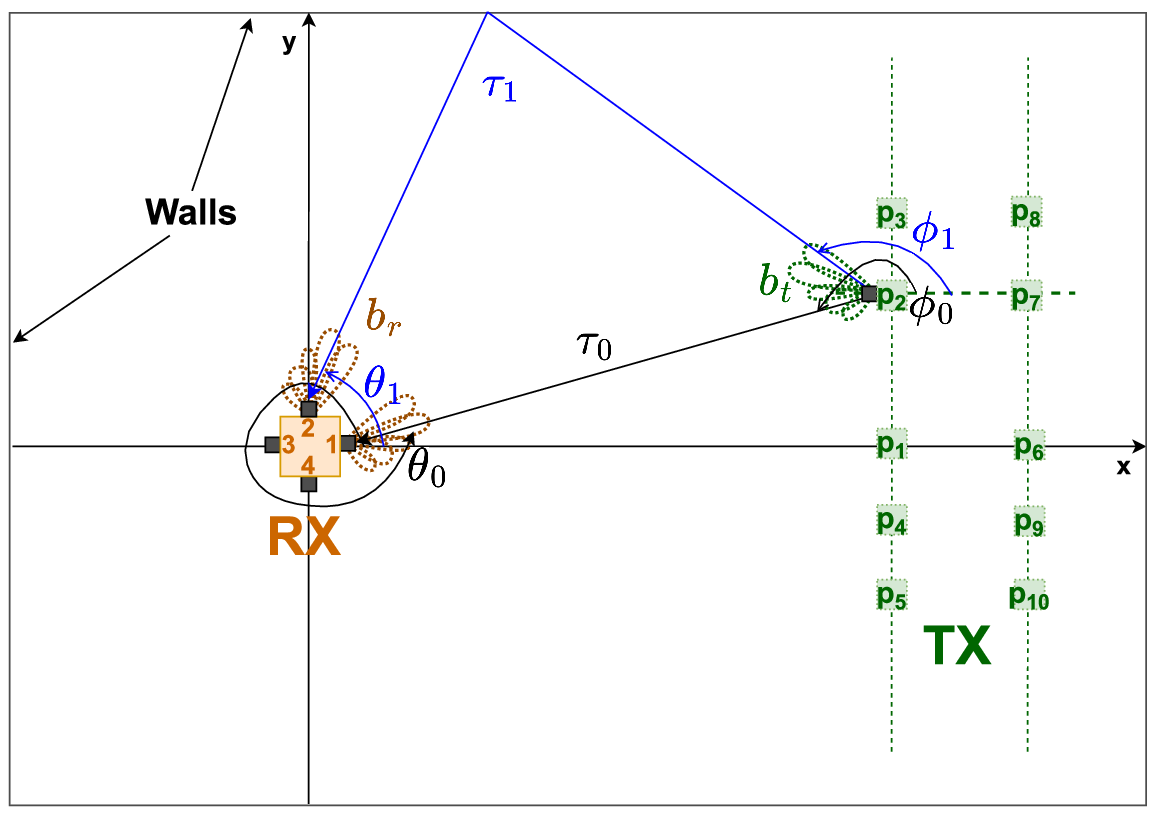}
    \caption{Floor plan of the room with RX placed at the origin, $\mathbf{p}_{1}$ to $\mathbf{p}_{10}$ represent the different TX static positions.}
    \label{layout}
\end{figure}

The position of the RX is stationary and known, while the TX is moved to different positions as shown in Fig.~\ref{layout}. We place the receiver's \lq center\rq \ at the origin with each of the 4 receivers placed at $\mathbf{p}_{\mathrm{RX},1} = [0.25 \, \mathrm{m}, 0 \, \mathrm{m}]$, $\mathbf{p}_{\mathrm{RX},2} = [0 \, \mathrm{m}, 0.25 \, \mathrm{m}]$, $\mathbf{p}_{\mathrm{RX},3} = [-0.25 \, \mathrm{m}, 0 \, \mathrm{m}]$, $\mathbf{p}_{\mathrm{RX},4} = [0 \, \mathrm{m}, -0.25 \, \mathrm{m}]$. The TX is placed at 10 different positions, $\mathbf{p}_{1} = [2.4 \, \mathrm{m}, 0 \, \mathrm{m}]$, $\mathbf{p}_{2} = [2.4 \, \mathrm{m}, 0.6 \, \mathrm{m}]$, $\mathbf{p}_{3} = [2.4 \, \mathrm{m}, 1.2 \, \mathrm{m}]$, $\mathbf{p}_{4} = [2.4 \, \mathrm{m}, -0.3 \, \mathrm{m}]$, $\mathbf{p}_{5} = [2.4 \, \mathrm{m}, -0.6 \, \mathrm{m}]$,  $\mathbf{p}_{6} = [3 \, \mathrm{m}, 0 \, \mathrm{m}]$, $\mathbf{p}_{7} = [3 \, \mathrm{m}, 0.6 \, \mathrm{m}]$, $\mathbf{p}_{8} = [3 \, \mathrm{m}, 1.2 \, \mathrm{m}]$, $\mathbf{p}_{9} = [3 \, \mathrm{m}, -0.3 \, \mathrm{m}]$, $\mathbf{p}_{10} = [3 \, \mathrm{m}, -0.6 \, \mathrm{m}]$. The positions were chosen to study the effect on the received signal laterally as well as longitudinally. The receiver has 4 antenna arrays placed orthogonally, while the transmitter faces one of the receiver arrays as shown in Fig.~\ref{layout}. Thus, we ensure LoS between the TX and $\mathrm{RX}_1$ but also nonline-of-sight (NLoS) on the other three RX. We use the measured RSS to estimate the AoA/AoD in azimuth only. The angles are measured w.r.t the plane parallel to the ground and counter-clockwise for each receiver. The synthetic environment is based on the measurement environment.

\section{System Model}

Consider an indoor office scenario with a TX equipped with a Uniform Linear Array(ULA) and a quasi-omnidirectional RX connected to 4 identical ULAs placed perpendicular to each other. In terms of RF chains, this forms a SIMO setup, which is chosen to enhance the number of MPCs. The transmitted signal $x[k] = [x_1, x_2, \ldots, x_K]$ is a time-varying, repeating Zadoff-Chu sequence as mentioned above.
The received signal measured at $\mathrm{RX}_i$, $i=1,\dots,4$, is
\begin{equation}
    y_{i,t,r}[k] = \sum_{l=0}^{L-1} a_l b_{r}(\theta_l) b_{t}(\phi_l) x[k - \tau_l]+  \eta[k] \ ,
    \label{eq2}
\end{equation}

where $L$ is the number of MPCs, $a_l$ is the path loss with amplitude $\alpha$ and phase $\varphi \in [0,2\pi]$ and $b_{r}(\theta_l)$ and $b_{t}(\phi_l)$ are beampattern weights for receiver and transmitter, respectively, with AoA $\theta$ and AoD $\phi \in [0,360^\circ]$.
$\tau_l$ is the time delay and $\eta$ is the noise with distribution $\mathcal{N} (x; \mu, \mathrm{N_0})$, $t$ and $r$ correspond to the beampattern numbers at the TX and RX, respectively. As can be observed from Fig.~\ref{beampattern}, each beampattern contributes to every sample of $\theta$ and/or $\phi$.

\section{The RSS-based Angle Estimation}
We first compute the RSS at each $\mathrm{RX}_i$ as, $$\mathrm{RSS}^{(i)}_{t,r} = \sum_{k=1}^{K} y[k] y[k]^* \ .$$
A first rough estimate of the AoA/AoD can be obtained by observing the RSS in the beampattern domain. Since the beampatterns are packed quite closely to each other, merely observing the power received across beampatterns provides us with an estimate for the general direction of the strongest beam of the signal. AoA can be assigned to the dominating angle of the beampattern with the maximum gain.

\begin{figure}[t]
    \centering
    \includegraphics[width = 0.92\linewidth]{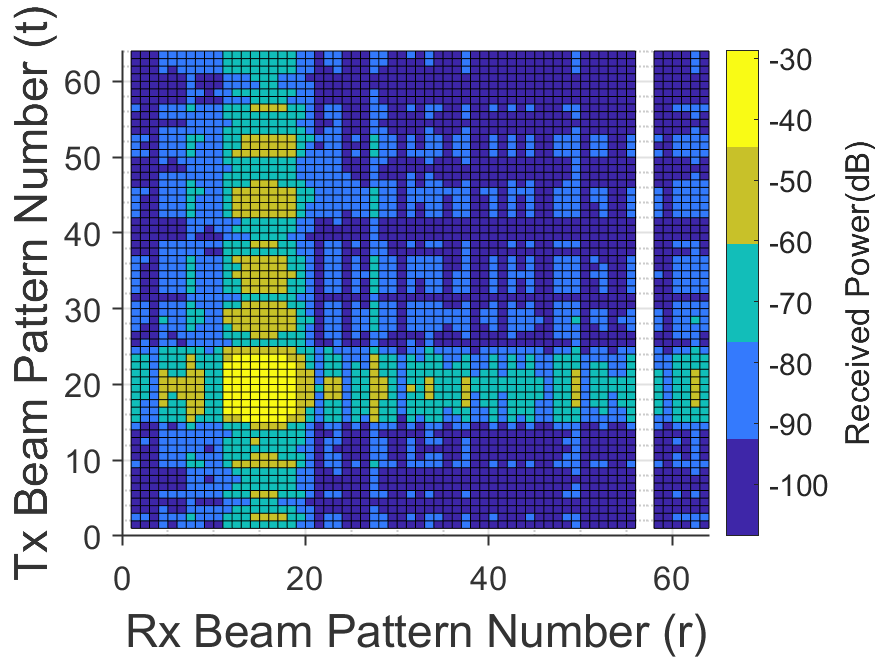}
    \caption{RSS plot for $\mathrm{RX}_1$ at position $\mathbf{p}_5$ showing received power across TX and RX beampatterns. Here we see the strongest power between beampatterns numbered $t = 15-25$ and $r = 10-20$.}
    \label{rss_p5}
\end{figure}

\begin{figure*}[t]
	\centering
	\subfloat[Receiver 1 (Measured)]{
		\includegraphics[scale=0.25]{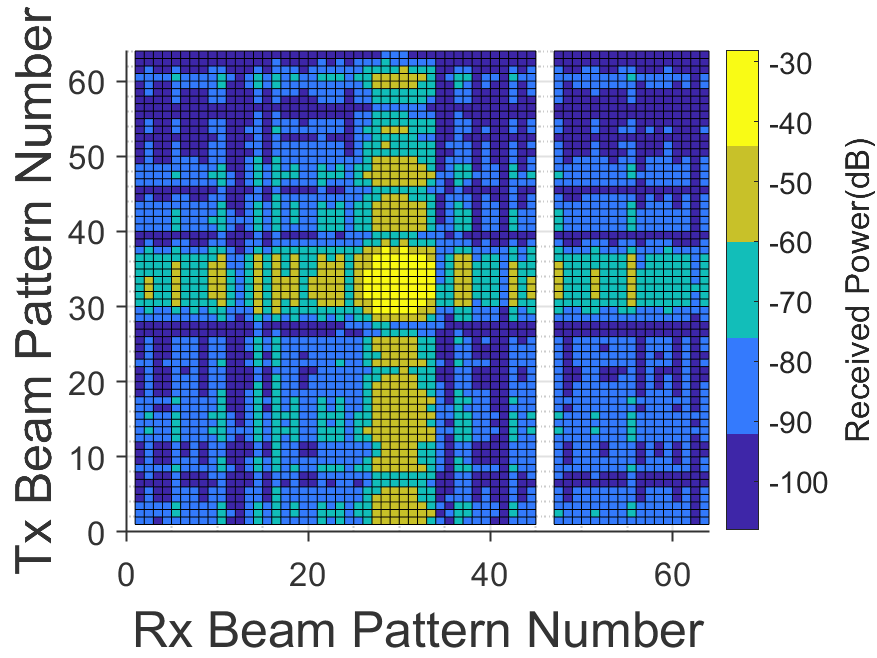}}
	~
	\subfloat[Receiver 1 (Simulated)]{
		\includegraphics[scale=0.25]{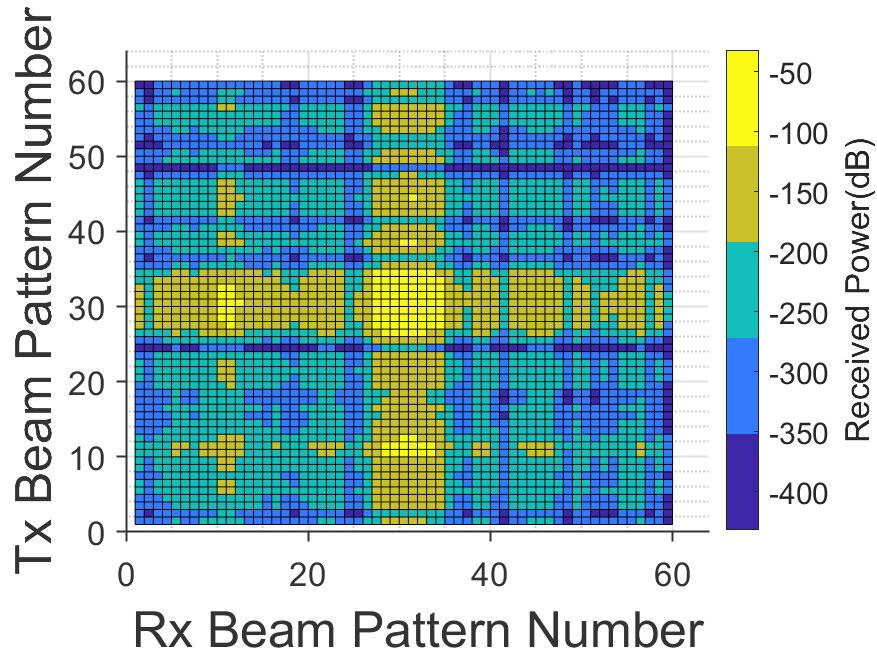}}
        ~
        \subfloat[Receiver 2 (Measured)]{
		\includegraphics[scale=0.25]{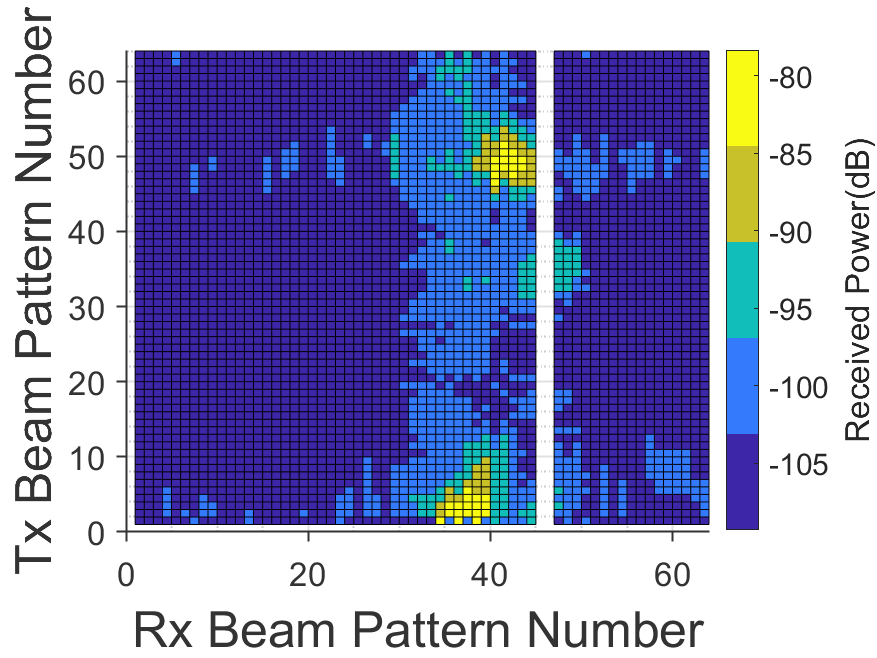}}
	~
	\subfloat[Receiver 2 (Simulated)]{
		\includegraphics[scale=0.25]{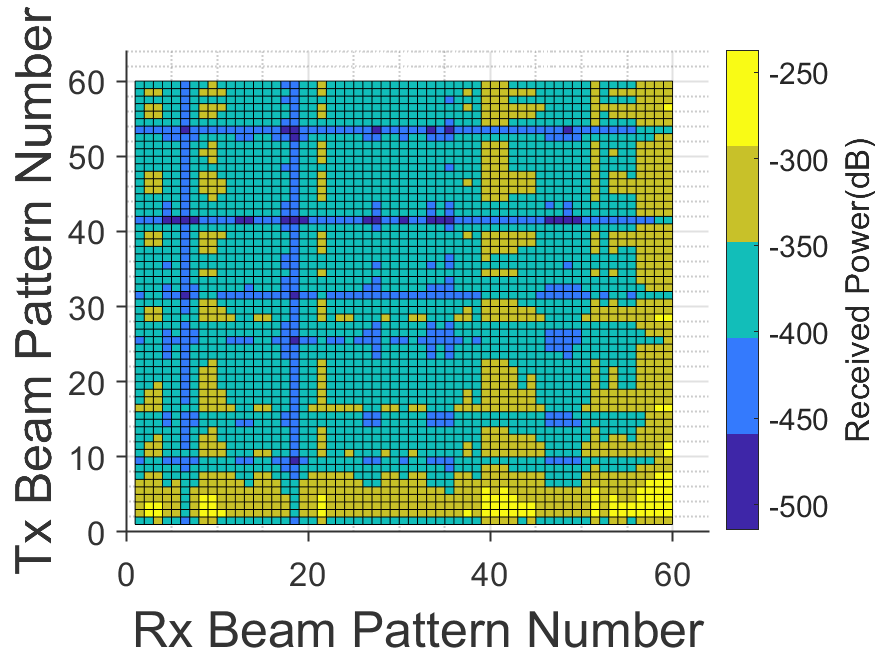}}
        \hfil

	\centering
	\subfloat[Receiver 3 (Measured)]{
		\includegraphics[scale=0.25]{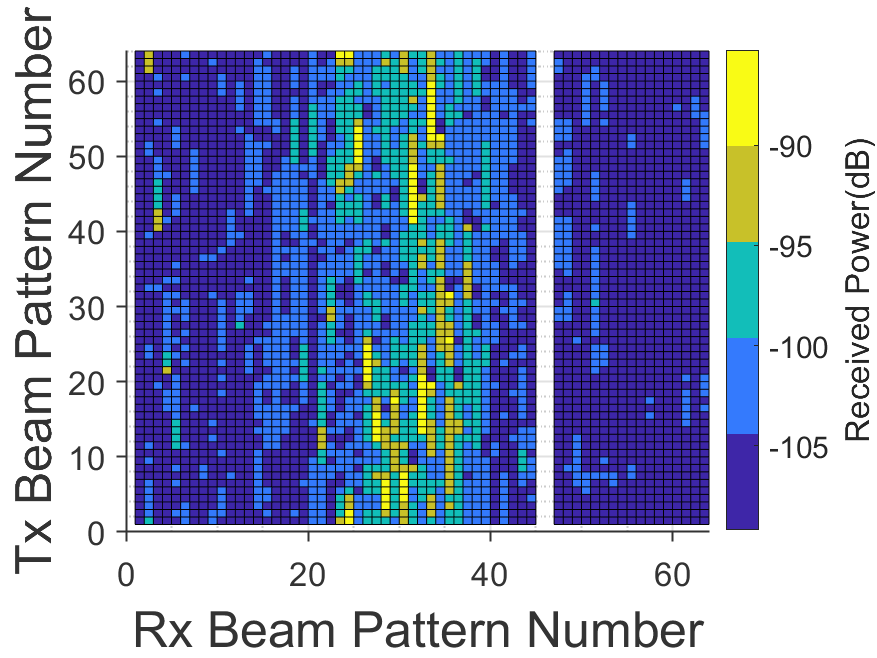}}
	~
	\subfloat[Receiver 3 (Simulated)]{
		\includegraphics[scale=0.25]{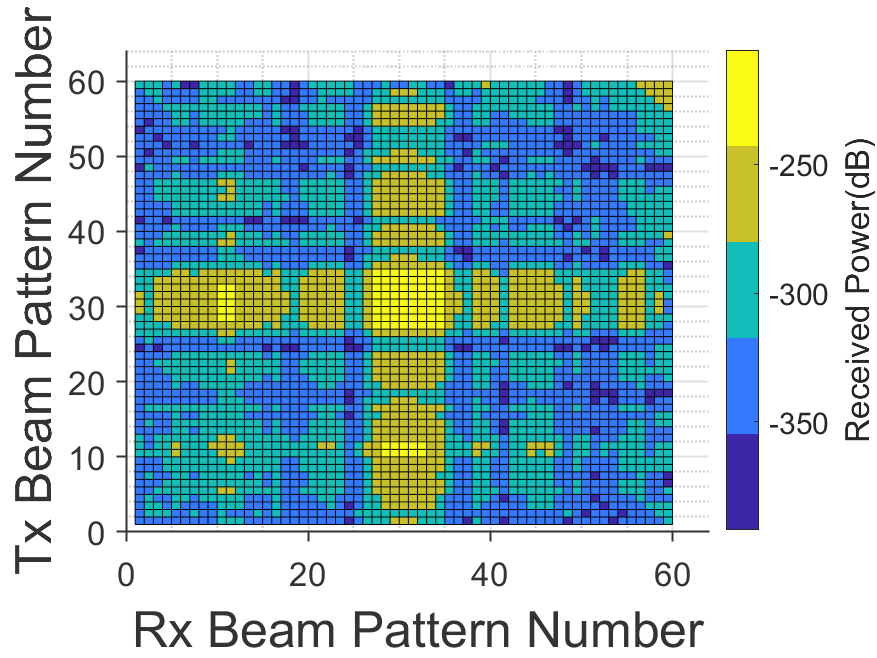}}
        ~
        \subfloat[Receiver 4 (Measured)]{
		\includegraphics[scale=0.25]{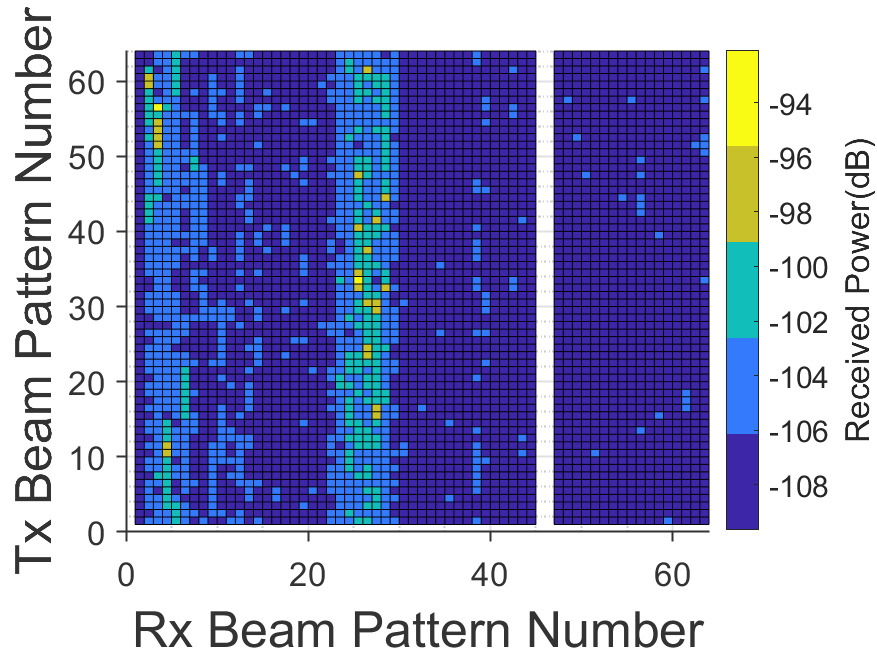}}
	~
	\subfloat[Receiver 4 (Simulated)]{
		\includegraphics[scale=0.25]{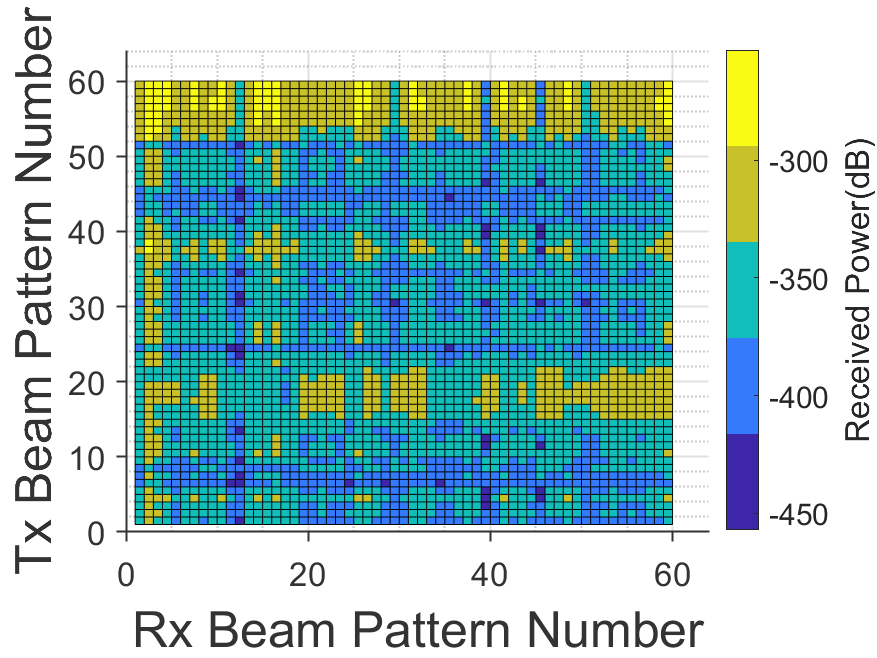}}
	\caption{Measured RSS data and simulated RSS data for all RX when TX is at position $\mathbf{p}_{1}$}
	\label{fig:result1}
\end{figure*}

\begin{figure*}[t]
	\centering
	\subfloat[Receiver 1 (Measured)]{
		\includegraphics[scale=0.25]{rx1/rss_POS5.eps}}
	~
	\subfloat[Receiver 1 (Simulated)]{
		\includegraphics[scale=0.25]{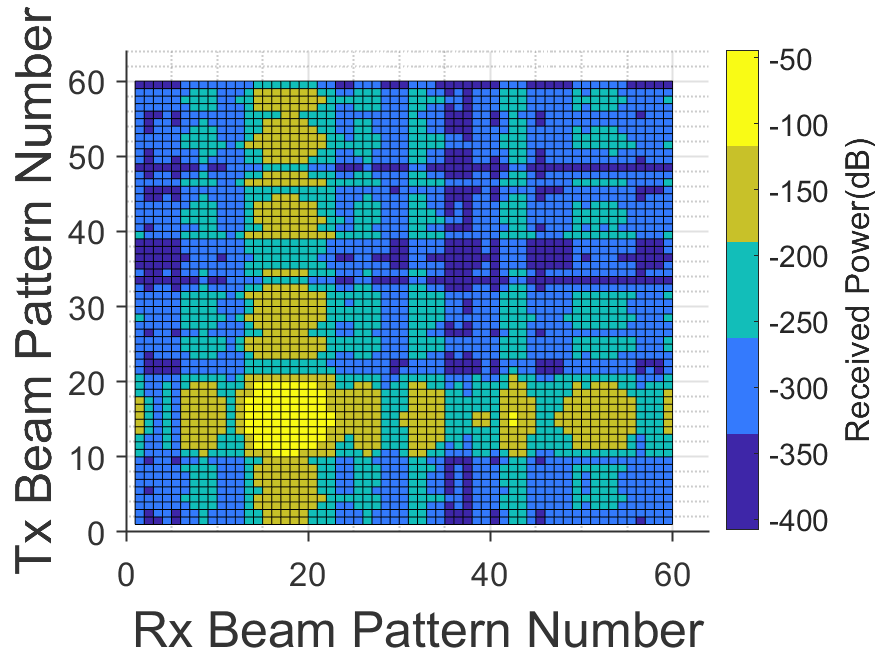}}
        ~
        \subfloat[Receiver 2 (Measured)]{
		\includegraphics[scale=0.25]{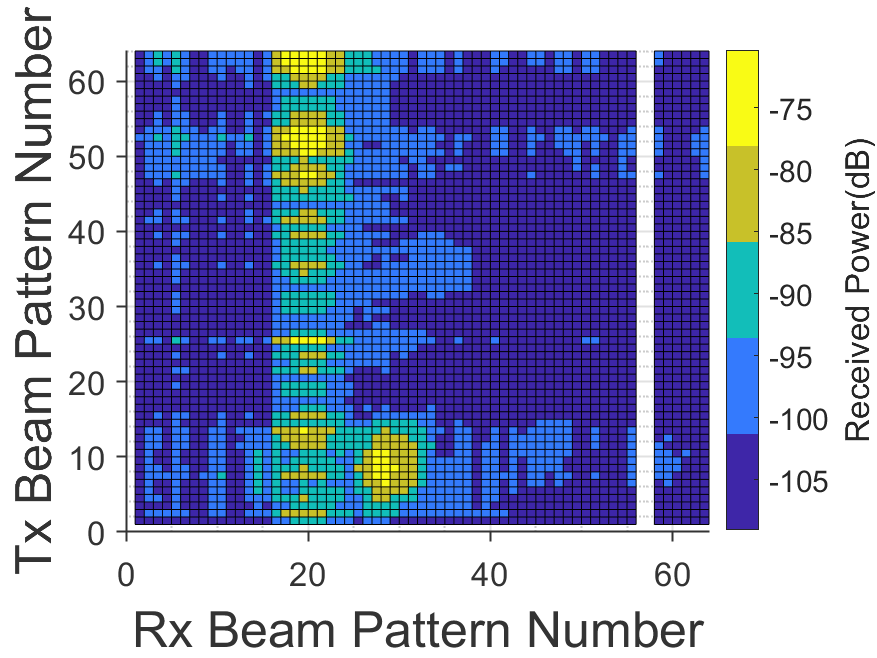}}
	~
	\subfloat[Receiver 2 (Simulated)]{
		\includegraphics[scale=0.25]{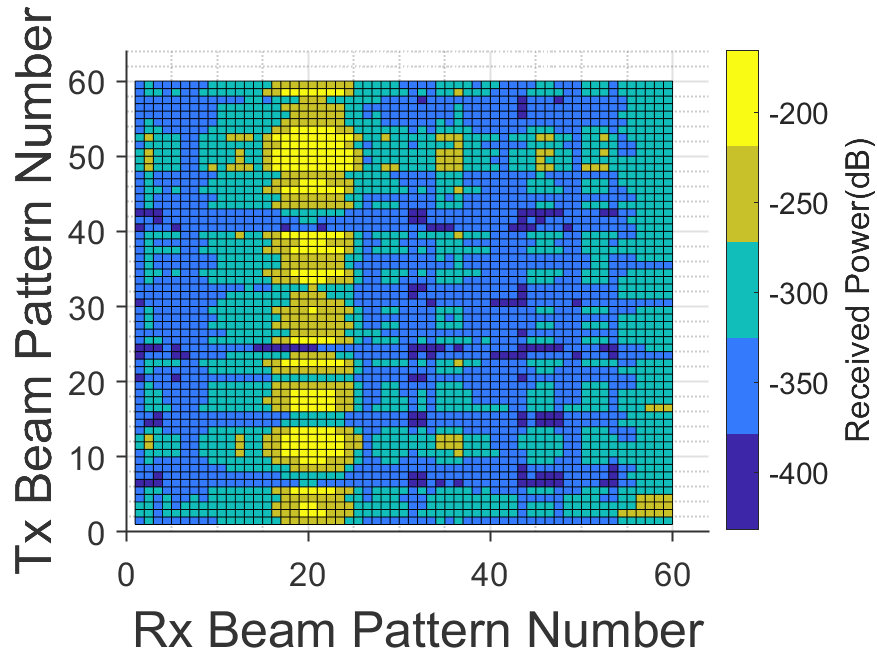}}
        \hfil

	\centering
	\subfloat[Receiver 3 (Measured)]{
		\includegraphics[scale=0.25]{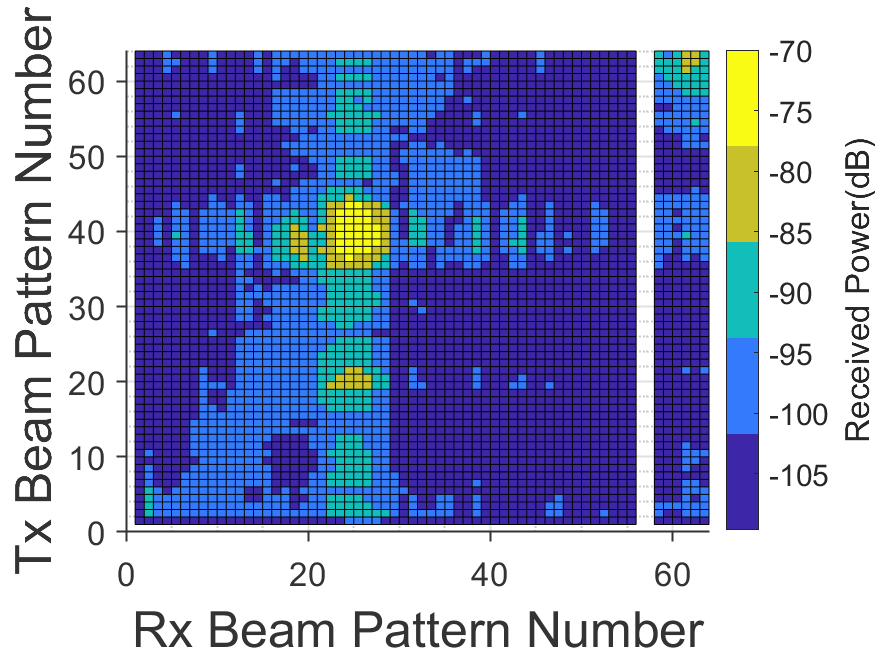}}
	~
	\subfloat[Receiver 3 (Simulated)]{
		\includegraphics[scale=0.25]{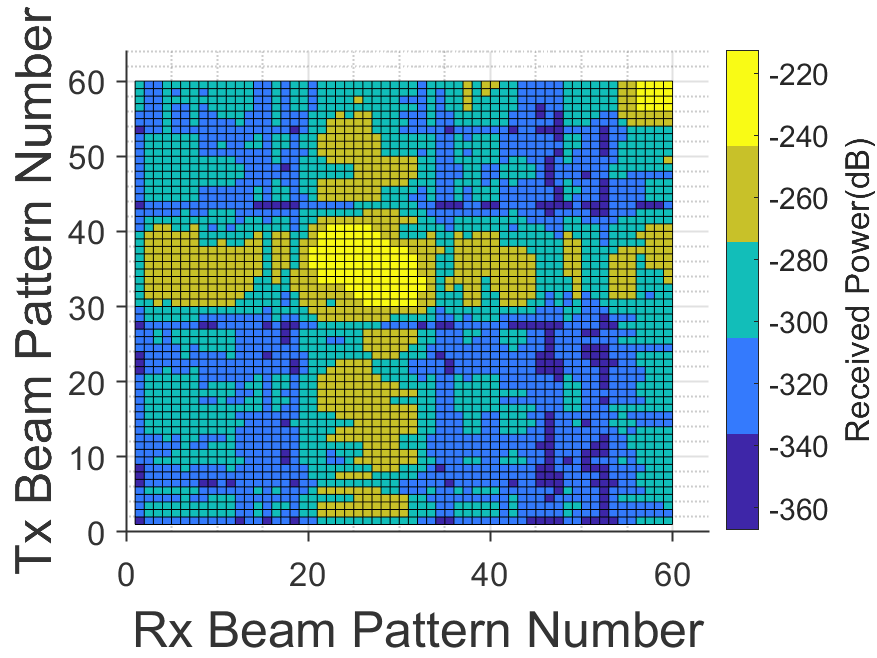}}
        ~
        \subfloat[Receiver 4 (Measured)]{
		\includegraphics[scale=0.25]{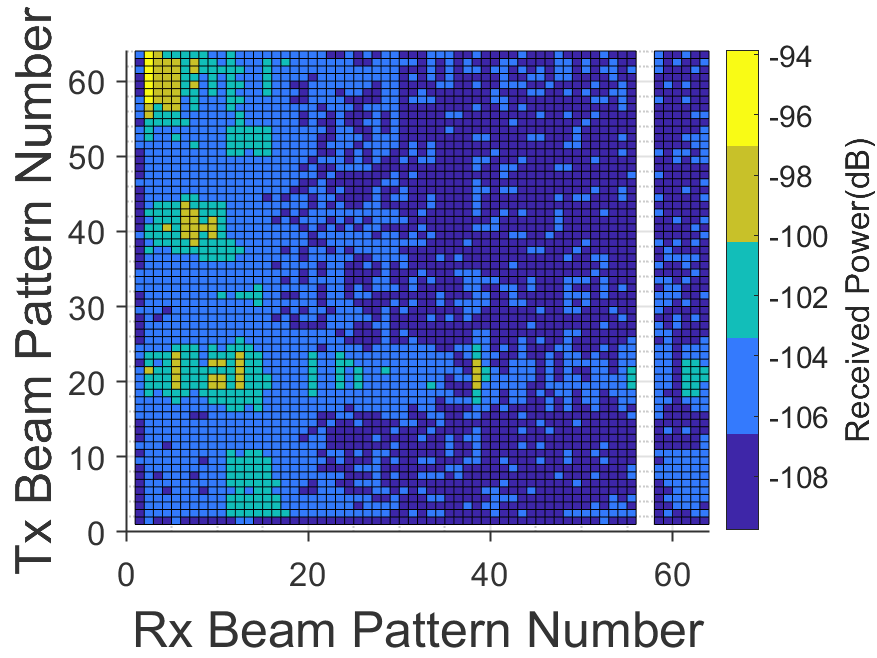}}
	~
	\subfloat[Receiver 4 (Simulated)]{
		\includegraphics[scale=0.25]{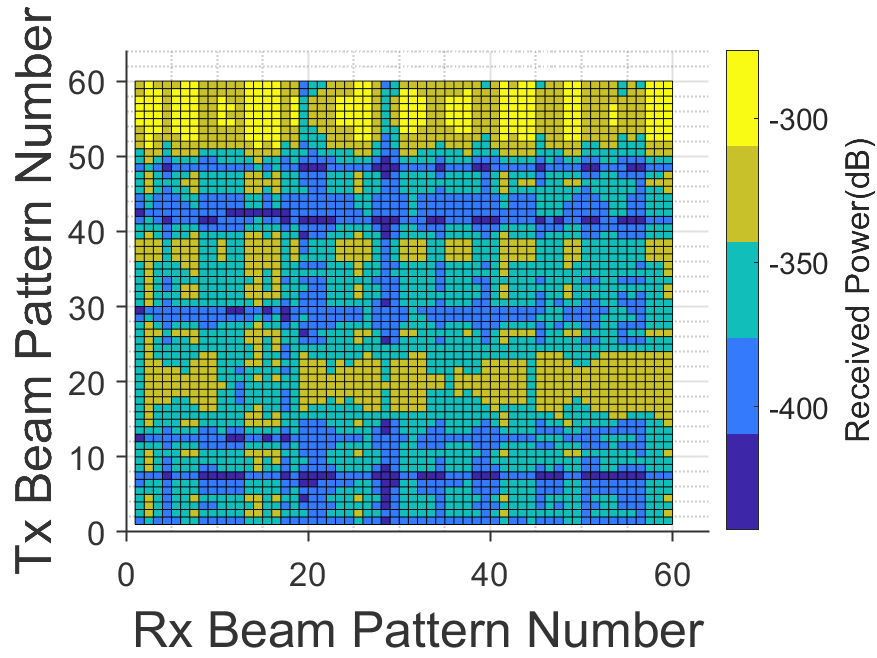}}
	\caption{Measured RSS data and simulated RSS data for all RX when TX is at position $\mathbf{p}_{5}$}
	\label{fig:result5}
\end{figure*}
An example of RSS is shown in Fig.~\ref{rss_p5}. Observing position $\mathbf{p}_{5}$ in Fig.~\ref{layout}, we see a LoS path between $\mathrm{TX}$ and $\mathrm{RX}_1$. The beampatterns move counterclockwise from 0 to 64 for both TX and RX, the aligning beampatterns should then be approx. 10-15 at TX and 15-20 at RX. This is seen in Fig.~\ref{rss_p5}, where the beampatterns with the strongest received power are between 10-20. We pick beampattern 14 with maximum RSS for TX and beampattern 19 at RX. Now, we find the angle where the beampattern has its main lobe and assign that as AoD/AoA which gives AoD as $-15^\circ$ and AoA as $-23.5^\circ$, respectively.

\subsection{RSS-LS2D Algorithm}

Let us now build a concrete algorithm that extends the single-angle estimation algorithm proposed in \cite{lacruz2020mm} to joint AoA/AoD estimation. Since our measurements bypass BRP, we obtain a full $N_\mathrm{\mathrm{BP},{TX}} \times N_\mathrm{\mathrm{BP},{RX}}$ beampattern grid across the TX and each of the four RX arrays. So, instead of just identifying one dominant beampattern, we formulate a least-squares problem utilizing the beampattern information at both the TX and RX. The joint AoD and AoA estimation can then be written as:
\begin{align}
   [\hat{\theta}, \hat{\phi}]  = & \argmin_{\theta, \phi} \min_{\alpha} \sum_{t,r}{}||{\mathrm{RSS}^{(i)}_{t,r} - \alpha b_r(\theta)b_t(\phi)}||^2 \\
%
%
 = & \argmin_{\theta, \phi} \min_{\alpha} \Big\{ |\alpha|^2  \sum_{t,r}|| b_r(\theta)b_t(\phi) ||^2 \label{eq3} \\ & - 2 \alpha \sum_{t,r} \mathrm{RSS}^{(i)}_{t,r} b_r(\theta)b_t(\phi) \Big\} \notag  
\end{align}

where $\alpha$ is the received signal amplitude for a pair of given beam patterns. Taking derivatives we find the $\alpha^{*}$ that minimizes the norm:

\begin{equation*}
 D_1(\alpha) = 2\alpha  \Big\{\sum_{t,r}|| b_r(\theta)b_t(\phi) ||^2\Big\} - 2 \sum_{t,r} \mathrm{RSS}^{(i)}_{t,r} b_r(\theta)b_t(\phi)
\end{equation*}

\begin{equation*}
 D_2(\alpha) =  2 \Big\{\sum_{t,r}|| b_r(\theta)b_t(\phi) ||^2\Big\} \geq 0 \ .
\end{equation*}

Here $D_1(\alpha)$ and $D_2(\alpha)$ are the first and second-order derivatives taken w.r.t $\alpha$.
Setting $ D_1(\alpha) = 0$ and substituting $\alpha^{*}$ in \eqref{eq3}, we obtain 

\begin{equation}
[\hat{\theta}, \hat{\phi}] =\argmin_{\theta, \phi} \left\{ -\frac{|\sum_{t,r} \mathrm{RSS}^{(i)}_{t,r} b_r(\theta)b_t(\phi)|^2}{\sum_{t,r}|| b_r(\theta)b_t(\phi) ||^2} \right\}
\end{equation}

\begin{equation}
[\hat{\theta}, \hat{\phi}] =\argmax_{\theta, \phi} \left\{ \frac{|\sum_{t,r} \mathrm{RSS}^{(i)}_{t,r} b_r(\theta)b_t(\phi)|^2}{\sum_{t,r}|| b_r(\theta)b_t(\phi) ||^2} \right\} \ .
\label{eq7}
\end{equation}

 To solve this two-dimensional problem, we run a grid search over all $\hat{\theta}$ and $\hat{\phi}$. For later discussion, with some abuse of notation we refer to $\mathbf{f(\theta,\phi)} = \frac{|\sum_{t,r} \mathrm{RSS}^{(i)}_{t,r} b_r(\theta)b_t(\phi)|^2}{\sum_{t,r}|| b_r(\theta)b_t(\phi) ||^2}$.

\subsection{RSS-LS1D Algorithm}
This algorithm is a variation of the algorithm mentioned in \cite{lacruz2020mm}, where the angle estimate is obtained by finding the strongest TX-RX beampattern pair using BRP and minimizing over a fixed beam. Since we do not use BRP, we go over each TX/RX beam pattern and minimize for $\theta$ or $\phi$ over the summation of all such correlations, respectively. 

\begin{equation}
   \hat{\theta} = \argmin_{\theta} \min_{\alpha} \sum_{t,r}{}||{\mathrm{RSS}^{(i)}_{t,r} - \alpha b_r(\theta)}||^2
   \label{eq9}
\end{equation}

\begin{equation}
   \hat{\phi} = \argmin_{\phi} \min_{\alpha} \sum_{t,r} ||{\mathrm{RSS}^{(i)}_{t,r} - \alpha b_t(\phi)}||^2
   \label{eq10}
\end{equation}

Similarly, solving for $\alpha$ in \eqref{eq9} and \eqref{eq10} we obtain:

\begin{equation}
\hat{\theta} =\argmax_{\theta} \left\{ {|\sum_{t} \mathrm{R}^{(i)}_{t} B_r(\theta)|^2} \right\}
\end{equation}

\begin{equation}
\hat{\phi} =\argmax_{\phi} \left\{ {|\sum_{r} \mathrm{R}^{(i)}_{r} B_t(\phi)|^2} \right\}
\end{equation}
where $\mathrm{R}^{(i)}_{t}$ and $\mathrm{R}^{(i)}_{r}$ are the normalized $\mathrm{RSS}^{(i)}_{t,r}$ values, such that $\mathrm{R}^{(i)}_{t} = \frac{\mathrm{RSS}^{(i)}_{t,r}}{||\mathrm{RSS}^{(i)}_{t} ||^2}$ and $\mathrm{R}^{(i)}_{r} = \frac{\mathrm{RSS}^{(i)}_{t,r}}{||\mathrm{RSS}^{(i)}_{r} ||^2}$. Analogously, $B_r(\theta)$ = $\frac{b_r(\theta)}{||b_r(\theta)||^2}$ and $B_t(\phi)$ = $\frac{b_t(\phi)}{||b_t(\phi)||^2}$ denote the normalized beampatterns.
Since this is a one-dimensional projection of the problem, the algorithm often locks to some local maxima instead of a global one as we will see in the next section. For later discussion, with some abuse of notation, we refer to $\mathbf{\zeta(\theta)} = {|\sum_{t} \mathrm{R}^{(i)}_{t} B_r(\theta)|^2}$ and  $\mathbf{\kappa(\phi)} = {|\sum_{r} \mathrm{R}^{(i)}_{r} B_t(\phi)|^2}$.

\section{Results and Discussion}
\label{sec:results}

The proposed algorithm is validated using measured and simulated data. The room at IMDEA is as shown in Fig.~\ref{setup3}. We created a 2D ray-tracing-based model assuming walls on all four sides. The windows, doors, and furniture were not modeled. First and second-order MPCs were obtained and applied to the signal model \eqref{eq2}. The model is used only for verification of measured results and therefore simplified. This generates I/Q sample data similar to what is obtained by the measurement setup. Each signal component (LoS and NLoS) experiences free space path loss and an additional 3 dB of attenuation per reflection for NLoS. The transmitted pulse is the same Zadoff-Chu sequence with $N_\mathrm{K} \times N_\mathrm{\mathrm{BP},{TX}}$ samples. For each component, the AWGN is generated having power spectral density $\mathrm{N_0}$, with $\mathrm{SNR} = 10 \log_{10}\frac{\lvert \lvert \alpha b_{r}(\theta_l) b_{t}(\phi_l) \rvert \rvert ^2}{\mathrm{N_0}}$, where $\alpha$ is the magnitude of the free space path loss, and $b_{r}(\theta_l)$ and $b_{t}(\phi_l)$ are the receive and transmit beampatterns, respectively.

The synthetic I/Q samples use the beampattern matrix obtained from measurements in an anechoic chamber (generated using the same codebook and antenna array as for measurements) and are of dimension  $N_\mathrm{\mathrm{BP},{TX}} \times N_\mathrm{\phi}$ and $N_\mathrm{\mathrm{BP},{RX}}\times N_\mathrm{\theta}$. Here $N_\mathrm{\mathrm{BP},{TX}} = N_\mathrm{\mathrm{BP},{RX}} = 64$ and $N_\mathrm{\theta}=N_\mathrm{\phi}$ are the angular samples ranging from $-79.5^\circ$ to $80^\circ$ with steps of $0.5^\circ$. The same measured beampattern data is used for angle estimation.

\begin{figure}
	\centering
	    \subfloat[AoD]{
		\adjincludegraphics[width =0.48\textwidth]{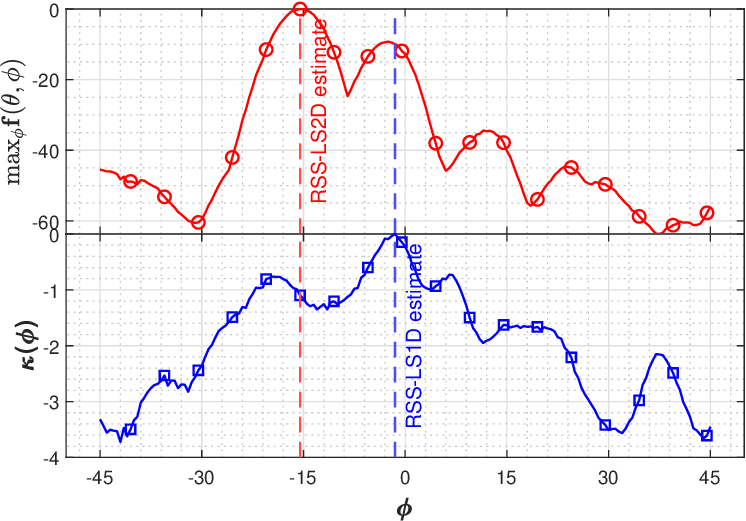}}
	\hfil
 
	\subfloat[AoA]{
		\includegraphics[width =0.48\textwidth]{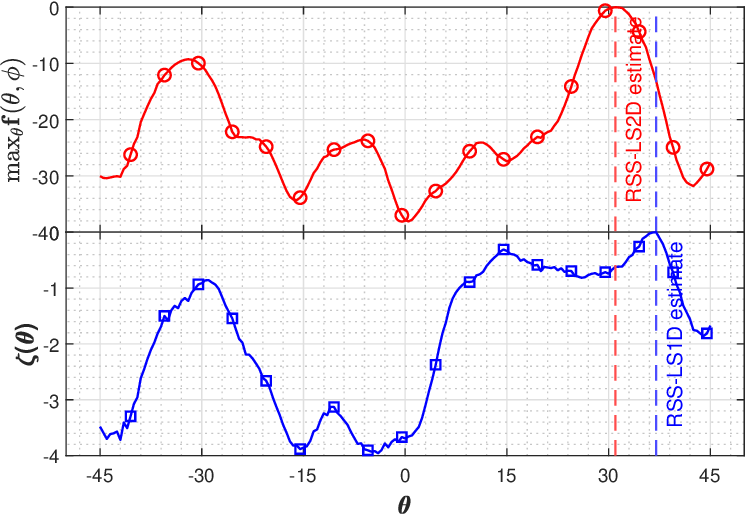}}

	\caption{Plots depicting the objective functions for $\mathbf{RX}_2$ when TX is at position $\mathbf{p}_5$. The two plots compare the performance of RSS-LS2D algorithm with the RSS-LS1D algorithm.}
	\label{fig:corr}
\end{figure}

\begin{table}[b]
\caption{Estimated AoA and AoD values for TX at $\mathbf{p}_1$}
\label{tab:pos1}
\centering
 \setlength{\tabcolsep}{2pt}
\begin{tabular}{|c|rrr|rrr|}
\hline
\multirow{2}{*}{} & \multicolumn{3}{c|}{AoA}                                                                              & \multicolumn{3}{c|}{AoD}                                                                               \\ \cline{2-7} 
                  & \multicolumn{1}{c|}{OR}           & \multicolumn{1}{c|}{RSS-LS2D}    & \multicolumn{1}{c|}{RSS-LS1D} & \multicolumn{1}{c|}{OR}           & \multicolumn{1}{c|}{RSS-LS2D}     & \multicolumn{1}{c|}{RSS-LS1D} \\ \hline
$\mathrm{RX}_1$   & \multicolumn{1}{r|}{$4^\circ$}     & \multicolumn{1}{r|}{$4^\circ$}   & $22.5^\circ$                    & \multicolumn{1}{r|}{$-2.5^\circ$}     & \multicolumn{1}{r|}{$-1^\circ$}   & $-3^\circ$                    \\ \hline
$\mathrm{RX}_2$   & \multicolumn{1}{r|}{$-41.5^\circ$}     & \multicolumn{1}{r|}{$-45^\circ$}  & $-29^\circ$                    & \multicolumn{1}{r|}{$9.5^\circ$}     & \multicolumn{1}{r|}{$10^\circ$}   & $-19.5^\circ$                    \\ \hline
$\mathrm{RX}_3$   & \multicolumn{1}{r|}{$22.5^\circ$}   & \multicolumn{1}{r|}{$21.5^\circ$} & $37^\circ$                  & \multicolumn{1}{r|}{$2^\circ$}     & \multicolumn{1}{r|}{$2.5^\circ$}  & $7^\circ$                  \\ \hline
$\mathrm{RX}_4$   & \multicolumn{1}{r|}{$5^\circ$} & \multicolumn{1}{r|}{$5.5^\circ$}  & $14.5^\circ$                  & \multicolumn{1}{r|}{$-7^\circ$} & \multicolumn{1}{r|}{$-6^\circ$} & $-3^\circ$                  \\ \hline
\end{tabular}
\end{table}
\begin{table}[t]
\caption{Estimated AoA and AoD values for TX at $\mathbf{p}_5$}
\label{tab:pos5}
\centering
 \setlength{\tabcolsep}{2pt}
\begin{tabular}{|c|rrr|rrr|}
\hline
\multirow{2}{*}{} & \multicolumn{3}{c|}{AoA}                                                                               & \multicolumn{3}{c|}{AoD}                                                                                 \\ \cline{2-7} 
                  & \multicolumn{1}{c|}{OR}          & \multicolumn{1}{c|}{RSS-LS2D}      & \multicolumn{1}{c|}{RSS-LS1D} & \multicolumn{1}{c|}{OR}           & \multicolumn{1}{c|}{RSS-LS2D}       & \multicolumn{1}{c|}{RSS-LS1D} \\ \hline
$\mathrm{RX}_1$                         & \multicolumn{1}{r|}{$-15^\circ$}  & \multicolumn{1}{r|}{$-16.5^\circ$} & $37^\circ$                    & \multicolumn{1}{r|}{$-23.5^\circ$} & \multicolumn{1}{r|}{$-22^\circ$}    & $-3^\circ$                    \\ \hline
$\mathrm{RX}_2$                         & \multicolumn{1}{r|}{$32^\circ$}   & \multicolumn{1}{r|}{$31^\circ$}    & $37^\circ$                    & \multicolumn{1}{r|}{$-15^\circ$}   & \multicolumn{1}{r|}{$-15.5^\circ$}  & $-1.5^\circ$                  \\ \hline
$\mathrm{RX}_3$                         & \multicolumn{1}{r|}{$15.5^\circ$} & \multicolumn{1}{r|}{$13.5^\circ$}  & $35^\circ$                    & \multicolumn{1}{r|}{$-8.5^\circ$}  & \multicolumn{1}{r|}{$-8^\circ$}     & $27.5^\circ$                  \\ \hline
$\mathrm{RX}_4$                         & \multicolumn{1}{r|}{$45^\circ$}   & \multicolumn{1}{r|}{$45^\circ$}    & $-3^\circ$                    & \multicolumn{1}{r|}{$-41.5^\circ$} & \multicolumn{1}{r|}{$-40.5^\circ$} & $-1.5^\circ$                  \\ \hline
\end{tabular}
\end{table}

First, we compute the RSS matrix at each receiver for all TX positions. As a sample, we can see in Fig.~\ref{fig:result1} and  Fig.~\ref{fig:result5} the RSS plots for positions $\mathbf{p}_1$ and $\mathbf{p}_5$ for all four receivers in comparison with their simulated counterparts. Looking at each position in the room layout and the geometry of the room we can predict some of the strongest paths for each RX. Some positions of the TX perform better than others depending on which receiver we observe. $\mathrm{RX}_4$ is the one facing the windows and we observe that the data obtained is unreliable, our simulations perform better because we do not model windows in our synthetic setup. If we were to estimate the angle by observing the strongest RSS, the angle estimate obtained is not bad due to the resolution of the beampatterns themselves. Still, it is prone to some errors especially for receivers not facing the TX, as can be seen for $\mathrm{RX}_4$ in a lot of cases. The white lines in the measured RSS plots signify the missing samples of certain beampatterns that weren't strong enough to activate the RX.

The obtained angle estimates from the two algorithms for two sample positions along with a rough estimate based on the observed RSS(OR) value are as in Table~\ref{tab:pos1} and \ref{tab:pos5}. We see that our proposed RSS-LS2D algorithm provides a more accurate angle estimate based on the $\theta, \phi$ grid minimization compared to RSS-LS1D which tends to vary from the true value. This is because the RSS-LS1D algorithm locks on local maxima, if we were to fix the strongest TX/RX beampattern it would be able to obtain the correct estimate. We notice, that the angle obtained from  RSS-LS1D is close to AoA/AoD of one of the MPCs just not the strongest one.

To elaborate on the performance of RSS-LS1D we observe the objective functions $\mathbf{\zeta(\theta)}$ and $\mathbf{\kappa(\phi)}$ as from \eqref{eq9} and \eqref{eq10} for $\mathrm{RX}_2$ in Fig~\ref{fig:corr}. We compare the function to a slice taken from $\mathbf{f(\theta,\phi)}$ of the RSS-LS2D objective function from \eqref{eq7}, where $\theta$ and $\phi$ are maximized. We are essentially observing the objective functions over which the $\argmax$ is computed. For instance, looking at Fig.~\ref{fig:corr}(a) we see that the peak maximized by the RSS-LS1D algorithm is present as a local maximum in the RSS-LS2D algorithm objective function $\mathbf{f}(\theta,\phi)$. But where RSS-LS2D maximizes at $\phi = -15^\circ$, RSS-LS1D maximizes at $\phi = -1.5^\circ$ (also seen in Table~\ref{tab:pos5}). For Fig.~\ref{fig:corr}(b), on the other hand, we see a similar behavior of the objective functions and maximized values relatively close to each other. This is to emphasize that we can observe plenty of cases where the RSS-LS1D tends to pick an MPC instead of the strongest path since it doesn't perform a maximization over the complete RSS grid.

\begin{figure}
	\centering
	\subfloat[Receiver 1]{
		\includegraphics[width =0.23\textwidth]{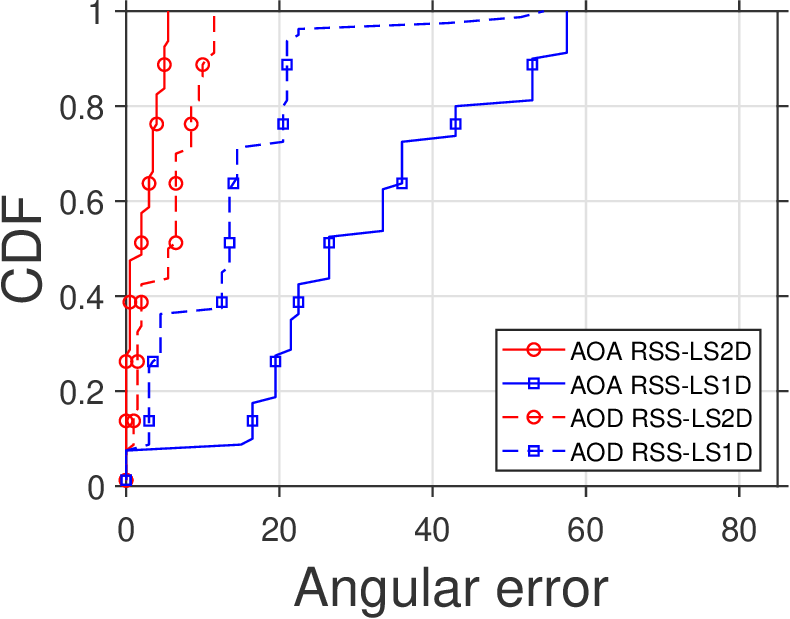}}
	~
	\subfloat[Receiver 2]{
		\includegraphics[width =0.23\textwidth]{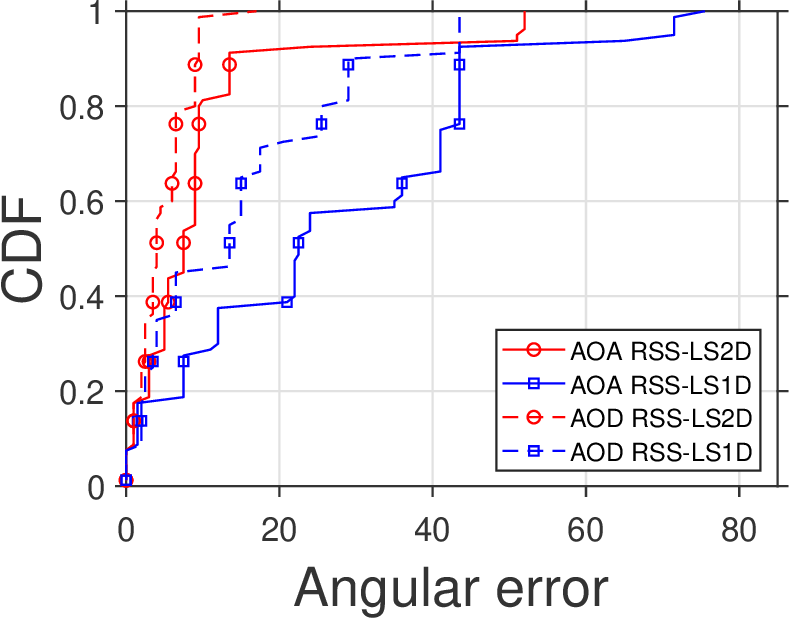}}
        \hfil

	\centering
	\subfloat[Receiver 3]{
		\includegraphics[width =0.23\textwidth]{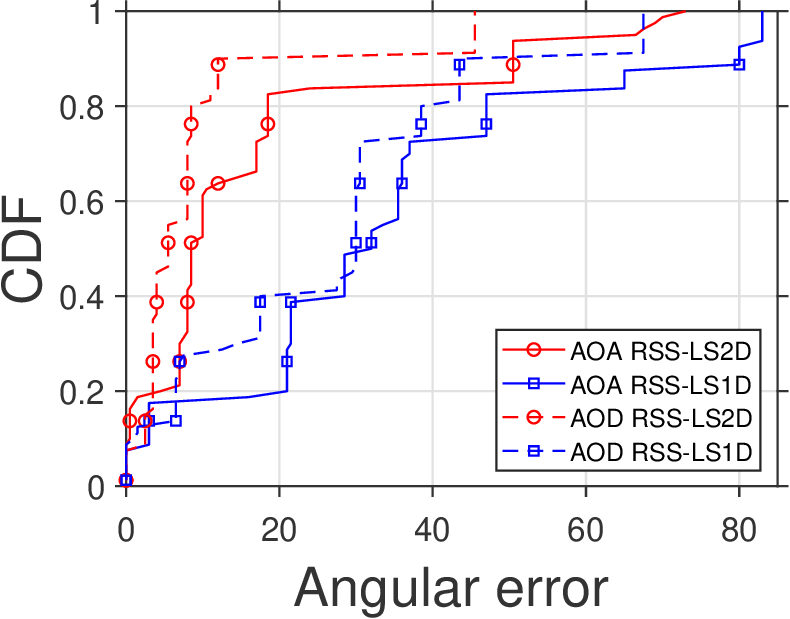}}
	~
	\subfloat[Receiver 4]{
		\includegraphics[width =0.23\textwidth]{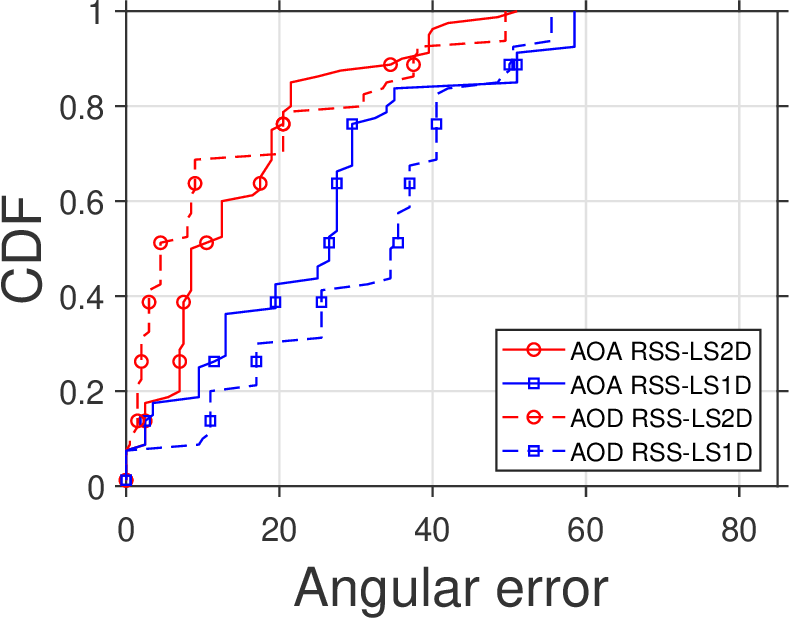}}
        
	\caption{CDF plots for error in AoA/AoD estimation using RSS-LS2D and RSS-LS1D over all positions.}
	\label{fig:cdf}
\end{figure}

To further compare the accuracy of our proposed algorithm, we compute the CDF of the error in AoA/AoD from the true value over all 10 receiver positions and 8 different snapshots of the measured data. The results are shown in Fig~\ref{fig:cdf}, where we can see that our proposed RSS-LS2D algorithm performs well, with less than a $10^\circ$ error in the majority of cases. 


\section{Conclusions}
\label{sec:conclusions}
In this paper, we show that even with mmWave directional signaling and only RSS information it is possible to exploit multipath information and obtain a good NLoS angle estimate. Our proposed algorithm RSS-LS2D can successfully estimate AoD and AoA for both LoS and NLoS cases with some stipulations. In particular, we outperform the RSS-LS1D which requires information about the strongest beampattern on at least one side to successfully estimate the angle on the opposite side. Although we only estimated one dominant path per RX, in the future we plan to estimate multiple AoA per RX array and add delay estimates to improve position accuracy.

\bibliographystyle{IEEEtran}
\bibliography{bibliography}

\end{document}